\documentclass[PRE,12pt]{revtex4}
\usepackage{natbib}
\usepackage{amssymb}
\usepackage{amsmath}
\usepackage{amsfonts}
\usepackage{graphicx}

\providecommand{\U}[1]{\protect\rule{.1in}{.1in}}

\begin{document}

\title{Thermodynamic Laws and Equipartition Theorem in Relativistic Brownian
Motion}
\author{T.~Koide}
\author{T.~Kodama}
\affiliation{Instituto de F\'{\i}sica, Universidade Federal do Rio de Janeiro, C.P.
68528, 21941-972, Rio de Janeiro, Brazil}

\begin{abstract}
We extend the stochastic energetics to a relativistic system. The
thermodynamic laws and equipartition theorem are discussed for a
relativistic Brownian particle and the first and the second law of
thermodynamics in this formalism are derived. The relation between the
relativistic equipartition relation and the rate of heat transfer is
discussed in the relativistic case together with the nature of the noise
term.
\end{abstract}

\maketitle

\section{Introduction}

\label{}

Physical interpretation and theoretical interpretation of Brownian motion, a
well-known process named after the Scottish botanist Robert Brown in 1827,
were first established by Einstein and Smoluchowski, then developed by
Langevin and many others. From these studies, it was made clear that
Brownian motion is nothing but the manifestation of the presence of
invisible microscopic bodies such as atoms and molecules, and their thermal
motion. Thus, Brownian motion has naturally been discussed in relation with
kinetic theories and thermodynamics. Nowadays concepts and methods of
stochastic processes introduced in the formulation of Brownian motion are
widely applied to various fields of science such as biophysics and economy.
Mathematical foundation of these methods has also been well studied. 

After thermodynamics was established in the nineteenth century, many efforts
have been made to understand the thermodynamic principles from the
microscopic point of view. The achievements of these efforts are summarized
today as statistical mechanics. There, the time average of microscopic
behaviors is replaced by an average in terms of a suitably chosen
statistical ensemble of microscopic events and the temporal behavior of each
particle is not questioned. Therefore, the basic question to the foundation
of statistical mechanics is how this substitution of averages is justified
from the microscopic deterministic point of view. Of course such a question
has been studied deeply in many ways since the time of Maxwell, Boltzmann
and Gibbs such as $H$-theorem and ergodic theory. Mathematically rigorous or
not, for physicists the overwhelming success of statistical mechanics itself
to describe the nature for thermally equilibrated systems could be
considered as the proof of validity of the hypothesis.

On the other hand, recently, it is more and more becoming important to
understand the bulk properties of a dynamical system where the thermodynamic
equilibrium is not necessarily attained. For example, in the field of
relativistic heavy ion collisions (RIC), we would like to determine the
thermodynamic properties of quark and gluon plasma (QGP) from the
experimental data. RIC is somehow the unique way in laboratories to extract
the thermodynamic properties of Quantum Chromodynamics (QCD) in the
deconfined phase. However, of course, real systems in RIC situation are not
in thermodynamic equilibrium, since the system size and the reaction time
scale are finite. Although many successful models, such as relativistic
hydrodynamics have been employed to describe several aspects of the observed
data, we feel still an existence of \textquotedblleft missing
link\textquotedblright\ between the thermodynamic properties and microscopic
dynamics. For example, several questions like, \textquotedblleft How much
can we be sure that the equation of state used to fit the experimental flow
data with hydrodynamics should be the same as that of obtained from the
lattice QCD calculations?\textquotedblright, \textquotedblleft Is the
extended thermodynamics valid and unique?\textquotedblright, or
\textquotedblleft How to deal with the relativistic covariance and
thermodynamic limit?\textquotedblright\ are yet difficult to be answered. In
this sense, it may still worthwhile to ask the repeated question for the
basic foundation of statistical mechanics from the view point of these new
physics necessities of nowadays.

The stochastic energetics may serve a useful tool to understand the
questions above \cite{sekimoto}. In this approach, the dynamics of each
event of a Brownian motion is directly reflected in the change of
thermodynamic quantities, such as heat, energy and work. That is, these
thermodynamic quantities are also dealt as stochastic variables, reflecting
in the name of this approach, \textit{stochastic energetics}. Therefore, the
application of this approach seems to be useful for RIC studies, where event
by event fluctuations are considered to carry important information on QCD 
\cite{ebe}.

So far, the stochastic energetics has been applied to non-relativistic
phenomena. However it seems very attractive to apply this approach to
investigate the thermodynamic properties for relativistic systems. This is
because it is possible to incorporate the effect of special relativity (such
as the Lorentz transform) into the model in a straight-forward manner,
permitting to discuss directly the relation between thermodynamics and the
theory of relativity. Such an analysis is difficult in statistical physics
where temporal averages are replaced by ensemble averages from the
beginning. The purpose of this paper is to show that the stochastic
energetics is still applicable to Brownian motion of a relativistic
particle. It is important to notice that this is not a trivial problem,
because equations of motion for stochastic variables involves a finite time
interval so that the space and time entangles when the Lorentz
transformation is involved. Therefore, we have to pay a special attention
for the treatment of noise when we discuss a relativistic stochastic motion
in a general reference frame.

The present paper is organized as follows: In Sec. 2, we introduce a model
of relativistic Brownian motion. Relativistic generalization of Brownian
motion has been discussed by many authors \cite%
{Dud,Hak,Ben,Oron,Boyer1,Boyer2,Pos,Deb,Fra,Hanggi1,Hanggi2,Zyga,rev_hanggi}%
, but the definitive answer to this problem has not yet established, or
simply does not exist in a unique manner. As mentioned above, this is
because of the difficulty to introduce a manifestly Lorentz covariant noise
in an unambiguous way, as well as its physical model. Here, instead of
formulating a covariant equation of motion for stochastic variables in a
general case, we discuss a less ambitious problem. We start from a Langevin
equation for relativistic Brownian motion defined in the rest frame of a
heat bath. Then we determine how the noise term should transform under an
arbitrary Lorentz boost of the system in order to keep the internal
consistency. We show that in such a situation, we can introduce a consistent
noise term. In Sec. 3, we apply the idea of the stochastic energetics to our
model of relativistic Brownian motion and derive the first and the second
law of thermodynamics. We also discuss the relation between the rate of heat
transfer and the relativistic equipartition relation. Summary and concluding
remarks are given in Sec. 4.

In this paper, we used the natural unit, $\hbar=c=k_{B} = 1$.

\section{Model of Relativistic Brownian Motion}

\label{sec2}

As mentioned in the introduction, the formulation of the relativistic
Langevin equation contains still an open question due to the transformation
property of the thermal noise. To avoid this question, we start with a
well-defined model of relativistic Brownian motion in the rest frame of a
heat bath. Then we study the transformation of the system under a Lorentz
boost and check the consistency between our model and special relativity.

\subsection{Relativistic Brownian motion in the rest frame of heat bath}

We consider the Brownian motion of a relativistic particle with mass $m$ in
the 3+1 dimension. In the rest frame of a heat bath, we consider the
following Langevin equation, 
\begin{subequations}
\begin{eqnarray}
\frac{d\mathbf{x}^{\ast }}{dt^{\ast }}&=& \frac{\mathbf{p}^{\ast }}{p^{0\ast
}},  \label{eqn:1} \\
\frac{d\mathbf{p}^{\ast }}{dt^{\ast }}&=& -\nu (p^{0\ast })\mathbf{p}^{\ast
}+\sqrt{2D(p^{0\ast })}\mathbf{N}(t^{\ast }),  \label{eqn:2}
\end{eqnarray}%
where $p^{0\ast }=\sqrt{(p^{\ast })^{2}+m^{2}}$, that is, the particle
always kept on the mass-shell. Throughout this paper, we use the symbol $%
^{\ast }$ to indicate the value of a variable defined in the rest frame of
the heat bath. The parameters $\nu (p^{0\ast })$ and $D(p^{0\ast })$
characterize, respectively, the relaxation of the momentum and the strength
of the noise. We assume that they are Lorentz scalar functions, depending
only on the particle energy in this frame. $\mathbf{N}$ is a Gaussian white
noise three-vector
\end{subequations}
\begin{equation}
\mathbf{N}(t^{\ast })=\left( 
\begin{array}{c}
N^{1} \\ 
N^{2} \\ 
N^{3}%
\end{array}%
\right) ,
\end{equation}%
and has the following correlation properties, 
\begin{align}
\langle \mathbf{N}(t^{\ast })\rangle _{\rm RF}& =0, \\
\langle N^{i}(t^{\ast })N^{j}(t^{\ast ^{\prime }})\rangle _{\rm RF}& =\delta
_{ij}\delta (t^{\ast }-t^{\ast ^{\prime }}).
\end{align}%
The symbol $\left\langle X\right\rangle _{\rm RF}$ denotes the stochastic
average of $X$ in the rest frame of the heat bath (we refer to as RF). The
similar Langevin equations have already discussed in Refs. \cite%
{Deb,Fra,Hanggi1,Hanggi2,Zyga,rev_hanggi}. Note that the Langevin equation
can be obtained even from a binary collision model \cite{Hanggi3}.

Now we replace the Langevin equation with the following stochastic
differential equation (SDE), 
\begin{subequations}
\begin{eqnarray}
d\mathbf{x}^{\ast} &=& \frac{\mathbf{p}^{\ast}}{p^{0\ast}}dt^{\ast}, \\
d\mathbf{p}^{\ast} &=& -\nu(p^{0\ast})\mathbf{p}^{\ast}dt^{\ast}+\sqrt{%
2D(p^{0\ast})}d\mathbf{w}_{t^{\ast}}.  \label{eqn:SDE1}
\end{eqnarray}
Here we have introduced the Wiener process $\mathbf{w}_{t^{\ast}}$ and its
difference by 
\end{subequations}
\begin{equation}
d\mathbf{w}_{t^{\ast}}\equiv\mathbf{w}_{t^{\ast}+dt^{\ast}}-\mathbf{w}%
_{t^{\ast}}=\mathbf{N}(t^{\ast})dt^{\ast}.
\end{equation}
The correlations are given by 
\begin{align}
\langle d\mathbf{w}_{t_{i}^{\ast}}^{\ast}\rangle_{\rm RF} & =0,
\label{eqn:corr1} \\
\langle d\mathbf{w}_{t_{k}^{\ast}}^{i\ast}d\mathbf{w}_{t_{l}^{\ast}}^{j\ast
}\rangle_{\rm RF} & =dt^{\ast}\delta_{ij}\delta_{kl}.  \label{eqn:corr2}
\end{align}

The last term of Eq. (\ref{eqn:SDE1}), 
\begin{equation}
\sqrt{2D(p^{0\ast})}d\mathbf{w}_{t^{\ast}}^{\ast}  \label{LastTerm}
\end{equation}
is a kind of Stieltjes integral for the stochastic variable. The definition
of Stieltjes integral in stochastic variables is known to be not unique.
Here we consider the three typical cases. Note that this problem associated
with the discretization scheme is discussed in Refs. \cite{Hanggi1,Hanggi2}
in detail.

\begin{enumerate}
\item Ito interpretation \cite{handbook}:

In this case, the term (\ref{LastTerm}) is interpreted as 
\begin{equation}
\sqrt{2D(p^{0\ast})} \circledast d\mathbf{w}_{t^{\ast}}^{\ast}=\sqrt{2D(p^{0\ast
}(t^{\ast}))}(\mathbf{w}_{t^{\ast}+dt^{\ast}}^{\ast}-\mathbf{w}%
_{t^{\ast}}^{\ast}).
\end{equation}
Hereafter, we use the symbol $\circledast$ to indicate the Ito interpretation of
the integral measure.

\item Stratonovich-Fisk interpretation \cite{handbook}:

In this case (hereafter, we use the symbol $\circ$ ), the term (\ref%
{LastTerm}) is interpreted as 
\begin{align}
\sqrt{2D(p^{0\ast})}\circ d\mathbf{w}_{t^{\ast}}^{\ast} & = \frac {\sqrt{%
2D(p^{0\ast}(t^{\ast}+dt^{\ast}))}+\sqrt{2D(p^{0\ast}(t^{\ast})})}{2}(%
\mathbf{w}_{t^{\ast}+dt^{\ast}}^{\ast}-\mathbf{w}_{t^{\ast}}^{\ast })  \notag
\\
& = \sqrt{D(p^{0\ast})} \partial_{\mathbf{p}^{\ast}}\sqrt{D(p^{0\ast})}
dt^{\ast}+\sqrt{2D(p^{0\ast})} \circledast d\mathbf{w}_{t^{\ast}}^{\ast}.
\end{align}
where, in the second line, we used the Ito formula (See Appendix) to convert
the Stratonovich-Fisk interpretation by the Ito scheme.

\item H\"{a}nggi-Klimontovich \cite{Hanggi3,Hanggi4,Kli}:

In this case (hereafter, we use the symbol $\star$ ), the term (\ref%
{LastTerm}) as 
\begin{align}
\sqrt{2D(p^{0\ast})}\star dw_{t^{\ast}}^{\ast} & = \sqrt{2D(p^{0\ast
}(t^{\ast}+dt^{\ast}))}(\mathbf{w}_{t^{\ast}+dt^{\ast}}^{\ast}-\mathbf{w}%
_{t^{\ast}}^{\ast})  \notag \\
& = 2\sqrt{D(p^{0\ast})}(\partial_{\mathbf{p}^{\ast}}\sqrt{D(p^{0\ast})}%
)dt^{\ast}+\sqrt{2D(p^{0\ast})} \circledast d\mathbf{w}_{t^{\ast}}^{\ast} .
\end{align}
Here, again, we used the Ito formula to express this interpretation in terms
of Ito scheme. See Appendix.
\end{enumerate}

The equilibrium distribution function described by using these SDEs depends
on the integral schemes defined above. To see this, we introduce the
probability density in the phase space defined in the thermal bath rest
frame, 
\begin{equation}
\rho(\mathbf{x}^{\ast},\mathbf{p}^{\ast},t^{\ast})=\langle\delta ^{(3)}(%
\mathbf{x}-\mathbf{x}^{\ast}(t^{\ast}))\delta^{(3)}(\mathbf{p}-\mathbf{p}%
^{\ast}(t^{\ast}))\rangle_{\rm RF}.  \label{rho}
\end{equation}
The time evolution of $\rho(\mathbf{x}^{\ast},\mathbf{p}^{\ast},t^{\ast})$
is given by the Fokker-Planck equation in a unified way for the different
integral schemes as%
\begin{align}
\partial_{t^{\ast}}\rho(\mathbf{x}^{\ast},\mathbf{p}^{\ast},t^{\ast}) &
=-\sum_{i}\partial_{\mathbf{x}^{\ast}}^{i} ( \frac{p^{i\ast}}{p^{0\ast}}\rho(%
\mathbf{x}^{\ast},\mathbf{p}^{\ast},t^{\ast}) )+\sum_{i}\partial _{\mathbf{p}%
^{\ast}}^{i}(\nu(\mathbf{p}^{0\ast})p^{i\ast}\rho(\mathbf{x}^{\ast},\mathbf{p%
}^{\ast},t^{\ast}))  \notag \\
& +\sum_{i}\partial_{\mathbf{p}^{\ast}}^{i} ( D^{1-\alpha}(\mathbf{p}^{0\ast
})\partial_{\mathbf{p}^{\ast}}^{i}D^{\alpha}(\mathbf{p}^{0\ast})\rho (%
\mathbf{x}^{\ast},\mathbf{p}^{\ast},t^{\ast}) ).  \label{FP}
\end{align}
The values of the parameter $\alpha$ correspond to the different
discretization schemes: $\alpha=0$ for the H\"{a}nggi-Klimontovich scheme, $%
\alpha=1/2$ for the Stratonovich-Fisk scheme, and $\alpha=1$ for the Ito
scheme.

From this equation, we find that the corresponding equilibrium distribution
(spatially homogeneous and static) is given by 
\begin{equation}
\rho_{\rm eq}(\mathbf{x}^{\ast},\mathbf{p}^{\ast})\propto\exp{\ \left(
-\int^{p^{0}(\mathbf{p}^{\ast})}s\ ds\ \frac{\nu(s)}{D(s)}-\alpha\ln
D(p^{0}\left( \mathbf{p}^{\ast}\right) )\right) }.  \label{eqn:st1}
\end{equation}
Since the phase space volume element $d\Gamma=d^{3}\mathbf{x}^{*} d^{3}%
\mathbf{p}^{*} $ should be an invariant measure under the Lorentz
transformation of the reference frame, we conclude that this equilibrium
distribution (\ref{rho}) should be a Lorentz scalar function, although the
proof is not trivial at all \cite{Kampen, Deb2}. As we will see later, we
can define the transformation property of the noise by using this fact.

\subsection{Relativistic Brownian motion in a moving frame}

We consider the reference frame which is moving with a constant velocity $%
\mathbf{V}$ with respect to the rest frame of the heat bath. We refer to as 
\textit{MF-moving frame}. The four-momentum $dp^{\mu }$ in this frame is
then given by the Lorentz transform of $dp^{\ast \mu }$ as%
\begin{align}
dp^{\mu }& =\Lambda (\mathbf{V})dp^{\ast \mu }  \notag \\
& =\left( 
\begin{array}{cc}
\gamma (\mathbf{V}) & \beta (V)\mathbf{n}^{T}\gamma (\mathbf{V}) \\ 
\beta (V)\mathbf{n}\gamma (\mathbf{V}) & \gamma (\mathbf{V})P_{\parallel
}+Q_{\perp }%
\end{array}%
\right) \left( 
\begin{array}{c}
dp^{0\ast } \\ 
d\mathbf{p}^{\ast }%
\end{array}%
\right) ,  \label{dp_MF}
\end{align}%
where $\gamma (\mathbf{V})=1/\sqrt{1-\mathbf{V}^{2}}$ and $\beta (V)=|%
\mathbf{V}|$. The projection operators are defined by $P_{\parallel }=%
\mathbf{n}\mathbf{n}^{T}$ and $Q_{\perp }=1-P_{\parallel }$ with $\mathbf{n}=%
\mathbf{V}/|\mathbf{V}|$. In the proceeding calculations, we assume that the
particle is always on mass-shell $p^{0}=\sqrt{\mathbf{p}^{2}+m^{2}}$ during
the whole stochastic process so that the stochastic variables, $dp^{0\ast }$
and $d\mathbf{p}^{\ast }$ are not independent. We then have,%
\begin{align}
dp^{0\ast }=d\sqrt{\mathbf{p^{\ast }}^{2}+m^{2}}& =\left\{ \left( -\nu
(p^{0\ast })+(1-\alpha )\frac{D^{\prime }(p^{0\ast })}{p^{0\ast }}\right) 
\frac{(\mathbf{p}^{\ast })^{2}}{p^{0\ast }}\right.  \notag \\
& \left. +D(p^{0\ast })\left( \frac{3}{p^{0\ast }}-\frac{(\mathbf{p}^{\ast
})^{2}}{(p^{0\ast })^{3}}\right) \right\} dt^{\ast }+\sqrt{2D(p^{0\ast })}%
\frac{\mathbf{p}^{\ast }}{p^{0\ast }} \circledast d\mathbf{w}_{t^{\ast }}^{\ast }.
\end{align}%
where $D^{\prime }(x)=dD(x)/dx$. Substituting this expression into Eq. (\ref%
{dp_MF}), we obtain the SDE in the MF as 
\begin{align}
d\mathbf{p}& =\left( -\frac{\nu (u^{\mu }p_{\mu })\gamma (V)(p^{0}-\beta
(V)p_{\rm V})}{p^{0}}+(1-\alpha )\frac{D^{\prime }(u^{\mu }p_{\mu })}{p^{0}}%
\right) \left\{ \mathbf{p}-\frac{\beta (V)m^{2}\mathbf{n}}{p^{0}-\beta
(V)p_{\rm V}}\right\} dt  \notag \\
& +\beta \gamma (V)\frac{D(u^{\mu }p_{\mu })}{p^{0}}\left( 2+\frac{m^{2}}{%
\gamma ^{2}(V)(p^{0}-\beta (V)p_{\rm V})^{2}}\right) \mathbf{n}dt+\widehat{%
\mathbf{B}} \circledast d\mathbf{w}_{t^{\ast }}^{\ast },  \label{eqn:SDEALT}
\end{align}%
where $p_{\rm V}=\mathbf{n}^{T}\mathbf{p}$ and 
\begin{equation}
\widehat{\mathbf{B}}=\sqrt{2D(u^{\mu }p_{\mu })}\frac{\gamma ^{-1}(V)}{%
p^{0}-\beta (V)p_{\rm V}}\left\{ p^{0}P_{\parallel }+\gamma (V)\left(
p^{0}-\beta (V)p_{\rm V}+\beta (V) ( \mathbf{n} \mathbf{p}^T ) \right)
Q_{\perp }\right\} .
\end{equation}%
%
Here $u^{\mu }$ is the four velocity $\left( \gamma ,\gamma \mathbf{V}%
\right) ,$ normalized as $u^{\mu }u_{\mu }=1$. The presence of a second rank
tensor $\widehat{\mathbf{B}}$ indicates that the noise is no longer
isotropic in the moving frame \cite{Matsas}.

The last term is not yet transformed because it contains the noise $d\mathbf{%
w}_{t^{\ast}}^{\ast}$ which is defined only in the RF. We first introduce
the stochastic noise which shows the property of the Gaussian white noise in
the MF as%
\begin{align}
\langle d\mathbf{w}_{t^{\ast}}\rangle_{\rm MF} & =0,  \label{Noise1} \\
\langle d\mathbf{w}_{t_{l}^{\ast}}^{i}d\mathbf{w}_{t_{m}^{\ast}}^{j}%
\rangle_{\rm MF} & =dt^{\ast}\delta_{ij}\delta_{lm},  \label{Noise2}
\end{align}
Here the symbol $\left\langle X\right\rangle _{MF}$ denotes the stochastic
average of $X$ in the MF. As was assumed in previous works \cite%
{Dud,Hak,Ben,Oron,Boyer1,Boyer2,Pos,Deb,Fra,Hanggi1,Hanggi2}, we may
consider that the correlations of the noise term are Lorentz invariant in
the rest frame of the particle, and we could write 
\begin{align}
\langle d\mathbf{w}_{t}\rangle_{\rm MF} & =\langle d\mathbf{w}_{t^{\ast}}^{\ast
}\rangle_{RF}=0, \\
\tilde{\gamma}(\mathbf{p})\langle d\mathbf{w}_{t_{k}}^{i}d\mathbf{w}%
_{t_{l}}^{j}\rangle_{\rm MF} & =\gamma(\mathbf{p}^{\ast})\langle d\mathbf{w}%
_{t_{k}^{\ast}}^{i\ast}d\mathbf{w}_{t_{l}^{\ast}}^{j\ast}\rangle_{\rm RF}=d\tau%
\delta_{ij}\delta_{kl},
\end{align}
where $d\tau$ is the proper time of the particle, $\gamma(\mathbf{p}^{\ast})$
and $\tilde{\gamma}(\mathbf{p})$ are the Lorentz factors of the particle in
the RF and in the MF, respectively. They are related through%
\begin{equation}
\tilde{\gamma}(\mathbf{p})=(\Lambda(\mathbf{V})\Lambda(-\mathbf{p}^{\ast
}))^{00}=\gamma(\mathbf{V})\gamma(\mathbf{p}^{\ast})(1-\beta(V)\frac {%
p_{\rm V}^{\ast}}{p^{0\ast}}),
\end{equation}
where, now $\gamma(\mathbf{V})$ is the Lorentz factor associated to the
Lorentz transformation from the RF to the MF. In this case, we conclude that
the transformation property of the noise is given by 
\begin{equation}
d\mathbf{w}_{t^{\ast}}^{\ast}=\sqrt{\frac{dt^{\ast}}{dt}}d\mathbf{w}_{t}=%
\sqrt{\gamma(\mathbf{p})\tilde{\gamma}^{-1}(\mathbf{p}^{\ast})}d\mathbf{w}%
_{t}=\gamma^{1/2}(V)\sqrt{\frac{p^{0}-\beta(V)p_{\rm V}}{p^{0}}}d\mathbf{w}_{t}.
\label{Trans1}
\end{equation}

Although very reasonable, a proof of the above argument is not obvious. The
reason is that the Stieltjes integral associated with the noise term is
defined on the time span $dt$, so that $d\mathbf{w}_{t^{\ast}}^{\ast}$ is
non-local in the time $t$. Thus the Lorentz transformation entangles with
the integration scheme in the order of $dt$. Then the noise term itself is
not necessarily covariant but can constitute a Lorentz vector only together
with the force term. That is, the force part and the stochastic part could
be mixed. This might be understood well from the argument of the kinetic
derivation of hydrodynamics. In the rest frame of fluids, the velocities of
molecules are completely random and all the kinetic energy of molecules are
replaced by internal thermodynamic quantities. On the other hand, when
fluids move, a part of velocities of molecules contributes to the collective
flow of fluids. In this case, the first order correlation calculated in the
MF of the noise term in the rest frame of the heat bath would not
necessarily vanish, 
\begin{equation}
\langle d\mathbf{w}_{t^{\ast}}^{\ast}\rangle_{\rm MF}\neq0,
\end{equation}
even if 
$\langle d\mathbf{w}_{t^{\ast}}^{\ast}\rangle_{\rm RF}=0$. 

Considering the finiteness of $dt,$ let us write, instead of Eq. (\ref%
{Trans1}), the following more general transformation property of the noise, 
\begin{align}
d\mathbf{w}_{t^{\ast}}^{\ast} & =\sqrt{\frac{dt^{\ast}}{dt}}d\mathbf{w}_{t}+%
\mathbf{C}_{\mathbf{p}}dt  \notag \\
& =\gamma^{1/2}(V)\sqrt{\frac{p^{0}-\beta(V)p_{\rm V}}{p^{0}}}d\mathbf{w}_{t}+%
\mathbf{C}_{\mathbf{p}}dt.  \label{trans2}
\end{align}
Here $\mathbf{C}_{\mathbf{p}}dt$ term entangles with the force term in the
MF, which should be separated from the pure stochastic part $d\mathbf{w}_{t}$%
, satisfying Eqs. (\ref{Noise1}) and (\ref{Noise2}). With this definition
and using the Ito formula, the Langevin equation is given by 
\begin{eqnarray}
d\mathbf{p} & =\left( -\frac{\nu(u^{\mu}p_{\mu})\gamma(V)(p^{0}-%
\beta(V)p_{\rm V})}{p^{0}}+(1-\alpha)\frac{D^{\prime}(u^{\mu}p_{\mu})}{p^{0}}%
\right) \left\{ \mathbf{p}-\frac{\beta(V)m^{2}\mathbf{n}}{p^{0}-\beta(V)p_{\rm V}%
}\right\} dt  \notag \\
& +\beta(V)\gamma(V)\frac{D(u^{\mu}p_{\mu})}{p^{0}}\left( 2+\frac{m^{2}}{%
\gamma^{2}(V)(p^{0}-\beta(V)p_{\rm V})^{2}}\right) \mathbf{n}dt+\widehat{\mathbf{%
B}}\ \mathbf{C}_{p}dt+\mathbf{\tilde{B}} \circledast d\mathbf{w}_{t},
\label{P-boosted}
\end{eqnarray}
where 
\begin{equation}
\tilde{\mathbf{B}}=\sqrt{\frac{\gamma(V)(p^{0}-\beta(V)p_{\rm V})}{p^{0}}}%
\widehat{\mathbf{B}}\mathbf{.}
\end{equation}
The last term in Eq. (\ref{P-boosted}) should be calculated according to the
Ito scheme. Note that the Ito formula to convert the SDE from one scheme to
the other must be used only after the transformation of the noise, Eq. (\ref%
{trans2}).

The corresponding Fokker-Plank equation for the SDE (\ref{P-boosted}) is 
\begin{equation}
\partial_{t}\rho=-\sum_{i}\partial_{x}^{i}\frac{p^{i}}{p^{0}}\rho+\sum
_{i}\partial_{p}^{i}\left[ -\mathbf{A}^{i}+\frac{1}{2}\partial_{\mathbf{p}%
}^{j}(\tilde{\mathbf{B}}\tilde{\mathbf{B}}^{T})^{ij}\right] \rho ,
\end{equation}
where 
\begin{align}
\mathbf{A} & =\left( -\frac{\nu(u^{\mu}p_{\mu})\gamma(V)(p^{0}-\beta(V)p_{\rm V})%
}{p^{0}}+(1-\alpha)\frac{D^{\prime}(u^{\mu}p_{\mu})}{p^{0}}\right) \left\{ 
\mathbf{p}-\frac{\beta(V)m^{2}\mathbf{n}}{p^{0}-\beta(V)p_{\rm V}}\right\} 
\notag \\
& +\beta\gamma(V)\frac{D(u^{\mu}p_{\mu})}{p^{0}}\left( 2+\frac{m^{2}}{%
\gamma^{2}(V)(p^{0}-\beta(V)p_{\rm V})^{2}}\right) \mathbf{n}+\widehat{\mathbf{B}%
}\ \mathbf{C}_{\mathbf{p}}.
\end{align}
Consequently, the equilibrium distribution function is given by the solution
of the following equation, 
\begin{equation}
\left[ -\mathbf{A}^{i}+\frac{1}{2}\partial_{\mathbf{p}}^{j}(\tilde {\mathbf{B%
}}\tilde{\mathbf{B}}^{T})^{ij}\right] \rho_{\rm eq}(\mathbf{x},\mathbf{p})=0,
\end{equation}
leading to the equilibrium distribution $\rho_{\rm eq}$ satisfying 
\begin{equation}
(\partial_{\mathbf{p}}^{i}\rho_{\rm eq})/\rho_{\rm eq}=2\sum_{jkl}(\tilde{\mathbf{B}}%
\tilde{\mathbf{B}}^{T})_{ij}^{-1}\left( \mathbf{A}^{j}-\frac{1}{2}\partial_{%
\mathbf{p}}^{k}\left( \tilde{\mathbf{B}}\tilde{\mathbf{B}}^{T}\right)
^{jk}\right) .  \label{DeriMV}
\end{equation}
Since $\rho_{eq}$ should be a scalar function, the above expression must
coincide with the Lorentz transform of the logarithmic derivative of the
equilibrium distribution function obtained in the RF. A lengthy but
straightforward calculation shows that $\rho_{eq}$\textbf{\ }can be the
Lorentz scalar 
only when%
\begin{equation}
\mathbf{C}_{\mathbf{p}}\equiv0.  \label{C}
\end{equation}
Thus, for the present model of a Brownian motion with the noise from a given
heat bath, we conclude that the noise term transforms separately from the
force term when we go to one reference frame to the other. That is, the
transformation of the noise (\ref{Trans1}) is consistent with the special
relativity, and the corresponding SDE in a MF should be given by Eq. (\ref%
{eqn:SDEALT}). When the background does not satisfy this condition, it is
not obvious if we can always assume $\mathbf{C}_{\mathbf{p}}\equiv0$ from
the beginning. Our discussion here is only for the thermal background is
homogeneous and static.

\subsection{Generalized fluctuation-dissipation relation}

So far, we have not specified the parameter of the SDE, but when we demand
that the equilibrium distribution function should be given by the J\"{u}%
ttner distribution in a MF, 
\begin{equation}
\rho_{\rm eq}=Const.\times e^{-u^{\mu}p_{\mu}/T},
\end{equation}
where $T$ denotes temperature. From this condition, the parameters of the SDE
satisfy the following relation, 
\begin{equation}
\nu(u^{\mu}p_{\mu})=\frac{1}{u^{\mu}p_{\mu}}\left( \frac{D(u^{\mu}p_{\mu})}{T%
}-\alpha D^{\prime}(u^{\mu}p_{\mu})\right) ,  \label{Einstein}
\end{equation}
which is the generalized Einstein's fluctuation-dissipation relation of
relativistic Brownian motion. Note that this relation depends on $\alpha$,
showing the discretization scheme dependence when $D$ is not a constant.

\section{Relativistic Stochastic Energetics}

Once we have established the SDE for a relativistic particle embedded in a
heat bath, we can discuss the thermodynamic property of the system defined
as the ensemble of these relativistic Brownian particles. For this purpose,
from now on, we discuss only in the rest frame of the heat bath, 
\begin{subequations}
\begin{eqnarray}
d\mathbf{x}^{\ast} &=& \frac{\mathbf{p}^{\ast}}{p^{0\ast}}dt^{\ast }, \\
d\mathbf{p}^{\ast} &=& -\nabla\phi\ dt^{\ast}-\nu(p^{0\ast})\mathbf{p}^{\ast
}dt^{\ast}+\sqrt{2D(p^{0\ast})} \circledast d\mathbf{w}_{t^{\ast}},
\label{eqn:rse_1}
\end{eqnarray}
where we have not assumed the generalized fluctuation-dissipation relation.
For the sake of the discussion of the first law of thermodynamics, we
introduced a scalar potential $\phi(\mathbf{x}^{\ast},t^{\ast})$ where the
explicit time dependence of $\phi$ represents the effect of some external
forces.

\subsection{The First Law of Thermodynamics}

Following Ref. \cite{sekimoto}, let us introduce heat as a stochastic
variable associated with Brownian motion. Among the three contributions of
the forces in Eq.(\ref{eqn:rse_1}), the last two terms represent the
interactions between a system and the heat bath. The first term represents
the mechanical force through the scalar potential $\phi$. Thus the work done
by the heat bath, that is, the heat transfer, $d^{\prime}Q$ to a Brownian
particle is defined as\footnote{%
We use the notation $d^{\prime}$ for quantities which are, in general, not
perfect differentials.} 
\end{subequations}
\begin{equation}
d^{\prime}Q=\left( -\nu(p^{0\ast})\mathbf{p}^{\ast}+\sqrt{2D(p^{0\ast})}\cdot%
\frac{d\mathbf{w}_{t^{\ast}}}{dt^{\ast}}\right) \circ d\mathbf{x}^{\ast }.
\label{dq1}
\end{equation}
Using Eq.(\ref{eqn:rse_1}) we can rewrite Eq.(\ref{dq1}) as 
\begin{align}
d^{\prime}Q & =\left( \frac{d\mathbf{p}^{\ast}}{dt^{\ast}}+\nabla
\phi\right) \circ d\mathbf{x}^{\ast}  \notag \\
& =d\left( p^{0\ast}+\phi\right) -\frac{\partial\phi}{\partial t^{\ast}}%
dt^{\ast}.  \label{1st_law}
\end{align}
Here, we identify the first term represents the change of the total energy
of a Brownian particle $dE$ including its potential energy. The quantity $%
\left( \partial\phi/\partial t^{\ast}\right) dt^{\ast}$ in the last term is
the change of the energy contained in the scalar potential $\phi$ due to the
change in some external parameters contained in $\phi$. That is, it can be
interpreted as the mechanical work $d^{\prime}W$ done by the external forces
to the system. Thus, rewriting Eq.(\ref{1st_law}) as 
\begin{equation}
d^{\prime}Q=dE-d^{\prime}W,  \label{1st}
\end{equation}
we can view this as the first law of thermodynamics in the framework of the
stochastic energetics, the generalization of Ref. \cite{sekimoto} for
relativistic Brownian motion. Note that this relation is satisfied for each
event of Brownian motion \footnote{%
Instead of introducing a scalar potential as in Eq.(\ref{eqn:rse_1}), we may
as well introduce it through the mass shift as $m\longrightarrow m-g\phi(%
\mathbf{x}^{\ast},t^{\ast})$, where $g$ is a coupling constant. This type of
interaction is known, for example, in the relativistic mean-field theory in
nuclear physics. Even if we use such a kind of potentials, it is still
possible to derive Eq. (\ref{1st}).}.

\subsection{The Second Law}

In the above, we introduced the potential $\phi$ to identify clearly the
role of external work in the first law of thermodynamics, and, there, the
choice of the parameters of the SDE was arbitrary. In the following
discussion of other thermodynamic relations, however, we will consider only
the case where $\phi=0$ and impose the generalized fluctuation-dissipation
theorem to the SDE.

For deriving the second law of thermodynamics in the stochastic energetics,
we introduce Shannon's information entropy, 
\begin{equation}
S=-\int d\Gamma\ \rho(\mathbf{x}^{\ast},\mathbf{p}^{\ast},t^{\ast})\ln \rho(%
\mathbf{x}^{\ast},\mathbf{p}^{\ast},t^{\ast}),  \label{Entro}
\end{equation}
where $\rho(\mathbf{x}^{\ast},\mathbf{p}^{\ast},t^{\ast})$ is defined in Eq.(%
\ref{rho}) and satisfies the Fokker-Plank equation (\ref{FP}). The
integration in Eq.(\ref{Entro}) is taken over all the phase space volume.
From these definitions for heat (\ref{dq1}) and entropy (\ref{Entro}), it is
easy to show that the time derivatives of heat and entropy are related as 
\begin{equation}
T\frac{dS}{dt^{\ast}}-\frac{1}{T}\left\langle \frac{d^{\prime}Q}{dt^{\ast}}%
\right\rangle _{\rm RF}=\int d\Gamma\frac{TD(p^{0\ast})}{\rho}\left( \nabla
_{p}\rho+\frac{1}{T}\frac{\mathbf{p}^{\ast}}{p^{0\ast}}\rho\right) ^{2},
\end{equation}
Since the right-hand side of the above equation is non-negative, we obtain
the following inequality, 
\begin{equation}
\frac{dS}{dt^{\ast}}-\frac{1}{T}\left\langle \frac{d^{\prime}Q}{dt^{\ast}}%
\right\rangle _{\rm RF}\geq0,
\end{equation}
which is nothing but the second law of thermodynamics. For the
non-relativistic case, see Ref. \cite{leb,sekimoto}. An important fact is
that, differently from the first law, the second law is satisfied only for
the average heat, and can be violated for each event, as is the case of the
fluctuation theorem.

\subsection{Relativistic Equipartition Theorem}

In the preceding section, we have shown that the first and the second laws of
thermodynamics can be consistently derived by generalizing the stochastic
energetics to relativistic Brownian motion. Let us apply this approach to
discuss another example, the equipartion theorem.

In non-relativistic systems, the equipartion theorem tells us that the
average kinetic energy of a particle of an ideal gas in equilibrium is
equally distributed by $T/2$ for each degree of freedom. For $3$-dimensional
mono-atomic gas, we have%
\begin{equation}
\left\langle \frac{\mathbf{p}^{2}}{2m}\right\rangle _{t\rightarrow\infty }=%
\frac{3}{2}T,  \label{nonrel_equi}
\end{equation}
where $\langle O\rangle_{t}$ denotes the expectation value of $O$ at the
instant $t,$ which can be expressed as%
\begin{equation}
\left\langle O\right\rangle _{t}=\int d\Gamma\ \rho\left( \mathbf{x,p,}%
t\right) O\left( \mathbf{x,p}\right) ,  \label{Av}
\end{equation}
with $\rho(\mathbf{x},\mathbf{p},t)$ is the solution of the Fokker-Plank
equation corresponding to the following non-relativistic SDE, 
\begin{subequations}
\begin{align}
d\mathbf{x} & =\frac{\mathbf{p}}{m}dt, \\
d\mathbf{p} & =-\nu\frac{\mathbf{p}}{m}dt-\nabla_{x}U(\mathbf{x})dt+\sqrt {2D%
} \circledast d\mathbf{w}_{t},
\end{align}
where $U(\mathbf{x})$ is a potential and $\nu=D/T$. Note that, the above
average (\ref{Av}) using the solution of the Fokker-Planck equation is
equivalent to the event average for the corresponding noise.

In the stochastic energetics, this relation for equipartition of energy is
seen to be valid in more general situation if the average heat transfer
vanishes even if the system is not in equilibrium \cite{sekimoto}. To see
this, we consider the case where $D$ is a constant so that the noise is
additive. Then, the time derivative of the heat described by this SDE is
given by 
\end{subequations}
\begin{align}
\left\langle \frac{d^{\prime}Q}{dt}\right\rangle & =-\frac{2D}{mT}\int
d\Gamma\left( \frac{\mathbf{p}^{2}}{2m}-\frac{3}{2}T\right) \rho (\mathbf{x},%
\mathbf{p},t)  \notag \\
& =-\frac{2D}{mT}\left( \left\langle \frac{\mathbf{p}^{2}}{2m}\right\rangle
_{t}-\frac{3}{2}T\right) ,  \label{NRheat}
\end{align}
where the left hand side average is the event average. One can see that the
equipartition relation is satisfied, when the heat transfer disappears even $%
\rho$ is not the equilibrium distribution. Thus we may consider the
condition of the null heat transfer leads to the equipartition relation in
more generalized situation. We will apply this idea to relativistic Brownian
motion.

The relativistic analogue is not trivial in the sense that there is no
simple interpretation of the quantity involved \cite{synge}. The expectation
value of relativistic kinetic energy, $K=\sqrt{p^{2}+m^{2}}-m$ does not
satisfy a simple relation such as Eq. (\ref{nonrel_equi}). Tolman proposed
to read Eq. (\ref{nonrel_equi}) as \cite{tolman,komer} 
\begin{equation}
\frac{1}{2}\left\langle \mathbf{p}\frac{\partial E}{\partial\mathbf{p}}%
\right\rangle _{\rm eq}=\frac{3}{2}T,  \label{Tolman}
\end{equation}
or equivalently, 
\begin{equation}
\left\langle \frac{\mathbf{p}^{2}}{2p^{0}}\right\rangle _{\rm eq}=\frac{3}{2}T,
\label{Tolman2}
\end{equation}
where the subscript $eq$ denotes the expectation value at equilibrium, that
is, $t\rightarrow\infty$. The left-hand side of Eq. (\ref{Tolman}) is
proportional to the average of $\mathbf{p\cdot v},$ which is interpreted to
be the momentum transfer rate than the kinetic energy itself.

As an application of the stochastic energetics for relativistic Brownian
motion, we can derive the relativistic analogue of Eq. (\ref{NRheat}).
Taking the noise average of Eq. (\ref{dq1}), we obtain 
\begin{align}
\left\langle \frac{d^{\prime}Q}{dt^{\ast}}\right\rangle _{\rm RF} &
=\left\langle \frac{d\mathbf{p}^{\ast}}{dt^{\ast}}\circ d\mathbf{x}%
^{\ast}\right\rangle _{\rm RF}=\left\langle d\mathbf{p}^{\ast}\circ\frac{d%
\mathbf{x}^{\ast}}{dt^{\ast}}\right\rangle _{\rm RF}  \notag \\
& =-\int d\Gamma\left\{ \frac{D(p^{0\ast})}{T}\frac{(\mathbf{p}^{\ast})^{2}}{%
(p^{0\ast})^{2}}-\left( \nabla_{p}\cdot\frac{D(p^{0\ast})\mathbf{p}^{\ast}}{%
p^{0\ast}}\right) \right\} \rho(\mathbf{x}^{\ast},\mathbf{p}^{\ast
},t^{\ast}) \nonumber \\
& =-\frac{1}{T}\left\langle D(p^{0\ast})\frac{(\mathbf{p}^{\ast})^{2}}{%
(p^{0\ast})^{2}}\right\rangle _{t^{\ast}}+\left\langle \nabla_{p}\cdot \frac{%
D(p^{0\ast})\mathbf{p}^{\ast}}{p^{0\ast}}\right\rangle _{t^{\ast}}.
\label{dq2}
\end{align}
To derive this expression, we have used the relation,%
\begin{equation}
\frac{\mathbf{p}^{\ast}}{p^{0\ast}}\circ d\mathbf{p}^{\ast}=d\mathbf{p}%
^{\ast } \circledast \left( \frac{\mathbf{p}^{\ast}}{p^{0\ast}}+\frac{1}{2}d\frac {%
\mathbf{p}^{\ast}}{p^{0\ast}}\right) .
\end{equation}
From Eq. (\ref{dq2}) the null heat transfer rate leads to a relation,%
\begin{equation}
\frac{1}{T}\left\langle D(p^{0\ast})\frac{(\mathbf{p}^{\ast})^{2}}{(p^{0\ast
})^{2}}\right\rangle _{t^{\ast}}=\left\langle \nabla_{p}\cdot\frac{%
D(p^{0\ast })\mathbf{p}^{\ast}}{p^{0\ast}}\right\rangle _{t^{\ast}},
\label{DQ0}
\end{equation}
independent of $\rho$. We call the above equation as the generalized
equipartition theorem for the relativistic Brownian motion. Note that this
relation does not have the same form of the Tolman relation (\ref{Tolman2}).
Only when the multiplicative noise is chosen as $D\propto p^{0\ast}$, Eq. (%
\ref{DQ0}) coincides with 
\begin{equation}
\left\langle \frac{\mathbf{p}^{2}}{2p^{0}}\right\rangle _{t^{\ast}}=\frac {3%
}{2}T,  \label{REQ}
\end{equation}
as in the non-relativistic case of Eq. (\ref{NRheat}) which is valid even in
out of equilibrium. In the ultra-relativistic limit, this relation is
reduced to%
\begin{equation}
\left\langle E\right\rangle _{t^{\ast}}=3T,
\end{equation}
where $E=p^{0\ast}$ so that it is related to the specific heat of the system.

The meaning of the equipartition theorem is modified for more general cases. For
example, suppose the noise has the energy dependence as $D\propto \left(
p^{0\ast }\right) ^{q}$ with $q=const$, then Eq.(\ref{DQ0}) becomes%
\begin{equation}
\left\langle E^{q}\right\rangle _{t^{\ast }}=T\times \left( q+2\right)
\left\langle E^{q-1}\right\rangle _{t^{\ast }},  \label{ept_ur}
\end{equation}%
in the ultra-relativistic limit, and 
\begin{equation}
\left\langle \left\vert \mathbf{p}\right\vert ^{q+2}\right\rangle _{t^{\ast
}}=Tm\times \left( q+3\right) \left\langle \left\vert \mathbf{p}\right\vert
^{q}\right\rangle _{t^{\ast }},  \label{epr_nonr}
\end{equation}%
in the non-relativistic Brownian motion (for non-relativistic case, Eq. (\ref%
{dq2}) is still the general form of Eq. (\ref{NRheat}) when the noise is
multiplicative, provided that $p^{0\ast }$ and $D(p^{0\ast })$ are $m$ and $%
D(\mathbf{p}^{2})$, respectively). These expressions show that the ratio of
two moments of kinetic energy should be kept constant proportional to the
temperature. In this sense we call Eq. (\ref{DQ0}) as the generalized
equipartition theorem.

For $q=-2$ in Eq. (\ref{ept_ur}) or $q=-3$ in Eq. (\ref{epr_nonr}), the left
hand side is finte, but the pre-factor in the right hand side vanishes. This
means that the average value in the right hand side in each equation
diverges but the products 
\begin{eqnarray}
&&\lim_{q\rightarrow -2}\left( q+2\right) \left\langle E^{q-1}\right\rangle
_{t^{\ast }},\  \\
&&\lim_{q\rightarrow -3}\left( q+3\right) \left\langle \left\vert \mathbf{p}%
\right\vert ^{q}\right\rangle _{t^{\ast }},
\end{eqnarray}%
have finite positive values and the equalities are still valid\textbf{. }%
This is easily checked for equilibrium cases, since we can calculate
explicitly 
\begin{eqnarray}
\langle E^{q}\rangle _{\rm eq} &=&\frac{4\pi }{Z}T^{q+3}\Gamma (q+3), \\
\langle \left\vert \mathbf{p}\right\vert ^{q+2}\rangle_{\rm eq}  &=&\frac{2\pi }{Z}%
(2mT)^{(q+5)/2}\Gamma ((q+5)/2),
\end{eqnarray}%
where $Z$ is the normalization factor of the equilibrium distribution
function and $\Gamma (x)$ is the Gamma function. From the recursion relation
of the Gamma function, we obtain Eqs. (\ref{ept_ur}) and (\ref{epr_nonr}).  Of
course these relations do not make sense for  $q<-2$ in Eq.(\ref{ept_ur}) or
for $q<-3$ in Eq.(\ref{epr_nonr}), since the left had side (so the right
hand side, too) diverges.

Note that the condition, Eq.(\ref{ept_ur}) or Eq.(\ref{epr_nonr}), itself
does not determine the distribution function. Of course, if we assume that
there exists a distribution function, $f(E;t^{\ast })$, independent of $q,$
satisfying Eq.(\ref{ept_ur}) for any $q>2,$ we can demonstrate that such a
distribution function should be proportional to $\exp (-E/T),$ and similary
for non-relativistic case. However, for a finite $t^{\ast },$ the
distribution function still reflects the initial condition and depends on $q.
$

Thus, as stated before, the null heat transfer condition does not mean that
the system is in equilibrium, but inversely, the null heat transfer
condition is naturally satisfied in thermal equilibrium for any choice of $%
D(p^{0\ast })$. For more general noise coefficient, this can be seen
directly from Eq. (\ref{dq2}) by re-expressing it as%
\begin{equation}
\left\langle \frac{d^{\prime }Q}{dt^{\ast }}\right\rangle _{\rm RF}=-\int
d\Gamma \left\{ \frac{D(p^{0\ast })}{T}\frac{(\mathbf{p}^{\ast })^{2}}{%
(p^{0\ast })^{2}}+\left( \frac{D(p^{0\ast })\mathbf{p}^{\ast }}{p^{0\ast }}%
\right) \cdot \left( \nabla _{p}\ln \rho (\mathbf{x}^{\ast },\mathbf{p}%
^{\ast },t^{\ast })\right) \right\} \rho (\mathbf{x}^{\ast },\mathbf{p}%
^{\ast },t^{\ast }).
\end{equation}%
The integrand vanishes if $\rho $ is given by the J\"{u}ttner distribution
function. Therefore, in thermal equilibrium, both of the generalized
equipartition theorem (\ref{DQ0}) and the Tolman relation (\ref{Tolman2})
are satisfied at the same time.

\begin{figure}[tbp]
\begin{minipage}[t]{6cm}
\begin{center}
\includegraphics[width=1.1\textwidth]{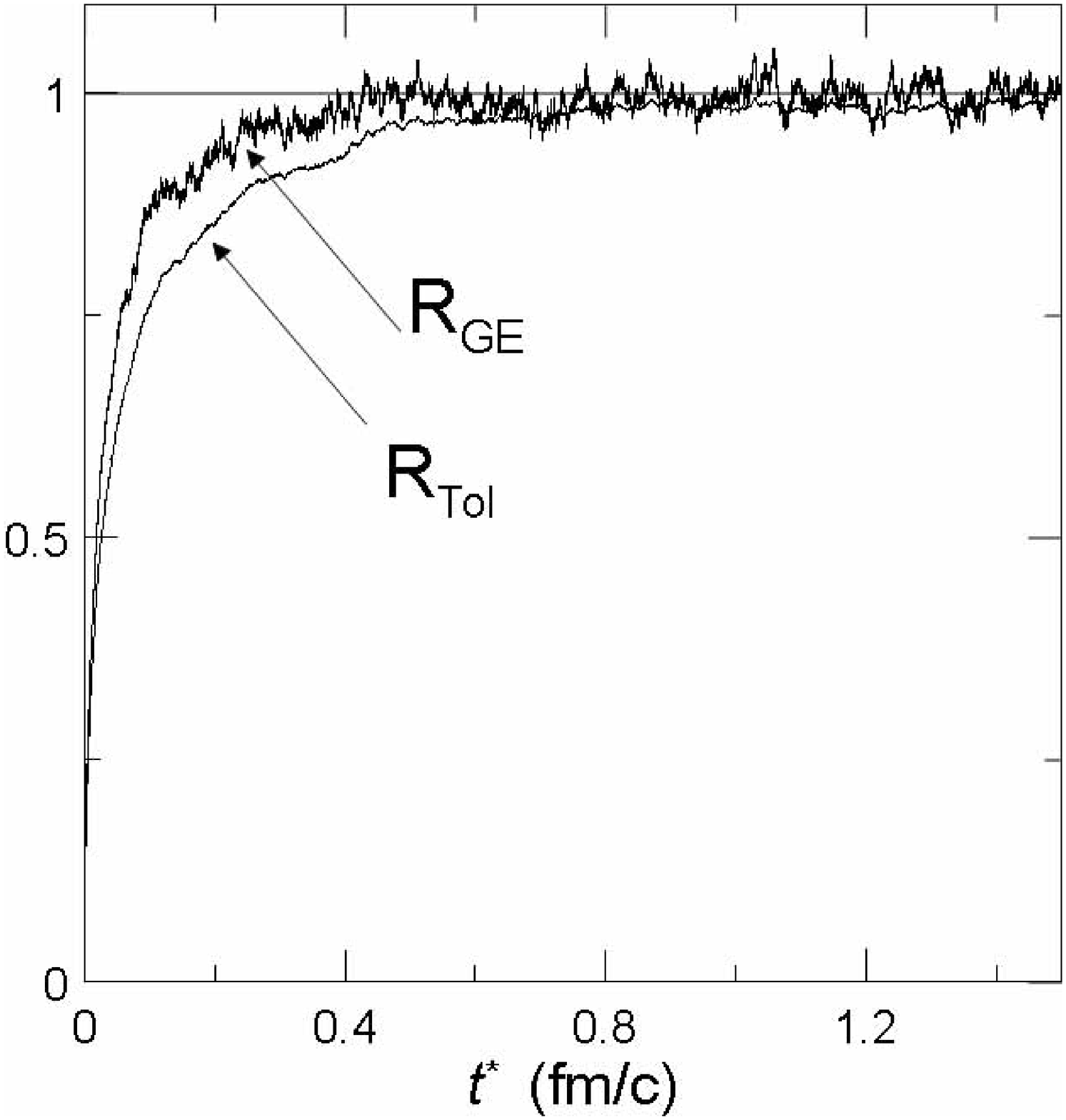}
\caption[Short caption for figure 1]{\label{labelFig1} The two ratios as a function of time $t^*$ with the initial condition of $50$ MeV.}
\label{fig1}
\end{center}
\end{minipage}
\hfill 
\begin{minipage}[t]{6cm}
\begin{center}
\includegraphics[width=1.1\textwidth]{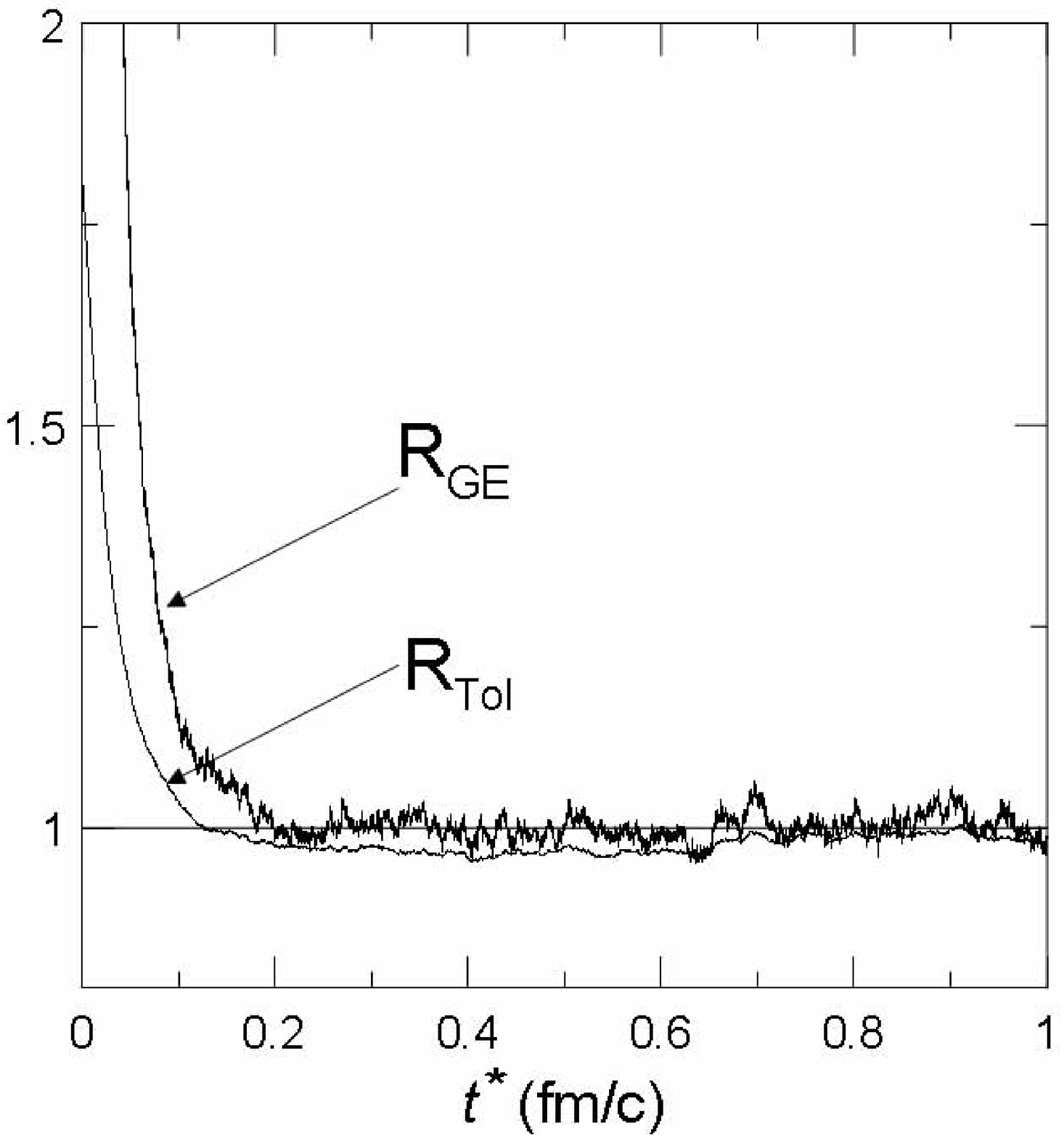}
\caption[Short caption for figure 2]{\label{labelFig2}  The two ratios as a function of time $t^*$ with the initial condition of $300$ MeV.}
\label{fig2}
\end{center}
\end{minipage}
\end{figure}

To see how the null heat transfer condition is attained, we consider the
case of a pion $m=139$ MeV in a heat bath of $T=150$ MeV as an example for a
possible application to RIC physics. In Figs. \ref{fig1} and \ref{fig2}, we
plot the ratio,%
\begin{equation}
R_{\rm GE}\left( t^{\ast }\right) =\frac{1}{T}\frac{\left\langle D(p^{0\ast })%
\frac{(\mathbf{p}^{\ast })^{2}}{(p^{0\ast })^{2}}\right\rangle _{t^{\ast }}}{%
\left\langle \nabla _{p}\cdot \frac{D(p^{0\ast })\mathbf{p}^{\ast }}{%
p^{0\ast }}\right\rangle _{t^{\ast }}},
\end{equation}%
by solving Eq. (\ref{eqn:rse_1}) for one dimensional case with $\phi =0.$ In
this example, we compare the two different situations; one is that the pion
is initially has a lower energy than the temperature of the heat bath, $%
p(t^{\ast}=0)=50$ MeV (Fig. \ref{fig1}), and the other the pion has
initially a higher energy, $p(t^{\ast }=0)=300$ MeV (Fig. \ref{fig2}). When
the null heat transfer $\left\langle d^{\prime }Q/dt^{\ast }\right\rangle
_{\rm RF}=0$ is attained, we should have $R_{\rm GE}\left( t^{\ast }\right) =1$. For
the sake of comparison, we show also the ratio 
\begin{equation}
R_{\rm Tol}\left( t^{\ast }\right) =\frac{1}{T}\left\langle \frac{p^{2}}{p^{0}}%
\right\rangle _{t^{\ast }},
\end{equation}%
which corresponds to the Tolman relation for one dimensional case. In
thermal equilibrium, this value should stay at $R_{\rm Tol}\left( t^{\ast
}\right) =1$. In these simulations, we consider the case of $q=-1,$ that is, 
$D(p^{0\ast })=K/p^{0\ast }$ with $K=0.1$ $\left( \mathrm{MeV}\right)^{4}$.
As seen from these figures, the null heat transfer condition is almost
satisfied before $t^{\ast }\simeq 0.5\ fm/c$, whereas the Tolman relation
only converges for $t^{\ast }>1$ $fm/c$. We checked some different parameter
sets and found that such a behavior always appears.

\section{Summary and Concluding remarks}

In this paper, we extended the stochastic energetics for relativistic
Brownian motion. As a model for the relativistic particle embedded in a heat
bath, we first established a relativistic SDE which produces the invariant J%
\"{u}ttner equilibrium distribution under an arbitrary Lorentz
transformation. Using this model, we discussed the thermodynamic laws and 
the equipartition theorem, by applying the stochastic energetics. We showed how
the first and second laws of thermodynamics are derived from relativistic
Brownian motion in this context. We obtained the explicit form of the heat
transfer rate between the relativistic Brownian particle and the heat bath.
Of course, our results recovers the corresponding results of Ref.\cite%
{sekimoto} in the non-relativistic limit. As a result, we showed that the
concept of the stochastic energetics is applicable to relativistic Brownian
motion.

We further showed that the condition of the null heat transfer leads to the
generalized equipartition theorem. Except for the particular choice of the
parameters of the SDE, the generalized equipartition theorem does not
coincide with Tolman's relativistic equipartition theorem if not in thermal
equilibrium. The null heat transfer and thermal equilibrium are not
equivalent, but the former includes the latter. As a matter of fact, we
found that the generalized equipartion relation is attained before the
system reaches the true thermal equilibrium in a few examples. It will be
interesting to investigate more in detail the meaning of null heat transfer
condition with respect to possible transient thermodynamic properties.

The stochastic energetics is considered as a promising approach for the
study of thermodynamics of mesoscopic systems \cite{sekimoto,
StochasticExample}. Thus, the present study will be a starting point of the
application of the stochastic energetics for the physics of relativistic
heavy ion collisions. There, the finite size of the system, as well as the
short reaction time are not negligible so that the deviation from the local
thermal equilibrium must be clarified. In particular, the approach via the
stochastic energetics allows introducing the energy conservation between
microscopic and macroscopic degrees of freedom. This may serve for a
consistent description of the formation of initial condition or the
freezeout processes of final states in the hydrodynamic modeling \cite%
{Hybrid}.

\hspace{2cm}

The authors acknowledge stimulating discussions with G. S. Denicol. This
work was (financially) supported by CNPq, FAPERJ, CAPES, PRONEX and the
Helmholtz International Center for FAIR within the framework of the LOEWE
program (Landesoffensive zur Entwicklung Wissenschaftlich- \"Okonomischer
Exzellenz) launched by the State of Hesse.


\appendix

\section{Ito formula}

\label{}

Let us consider an arbitrary function $f(\mathbf{x})$ and the evolution of $%
\mathbf{x}$ is given by the SDE, 
\begin{align}
d\mathbf{x}=\mathbf{A}dt+ \mathbf{B} \circledast d\mathbf{w}.
\end{align}
Then, the variation of $f(\mathbf{x})$ is 
\begin{equation}
df(\mathbf{x})=\{\sum_{i}\mathbf{A}^{i}\partial_{i}f(\mathbf{x})+\frac{1}{2}%
\sum_{ij}[\mathbf{B}\mathbf{B}^{T}]^{ij}\partial_{i}\partial_{j}f(\mathbf{x}%
)\}dt+\sum_{ij}\mathbf{B}^{ij}\partial_{i}f(\mathbf{x}) \circledast d\mathbf{w}^{j}.
\label{Ito}
\end{equation}
This is called the Ito formula \cite{handbook}. By using the Ito formula, we
obtain, 
\begin{align}
[\mathbf{B}\circ d\mathbf{w} ]^{i} & =[\mathbf{B} \circledast d\mathbf{w}]^{i}+%
\frac{1}{2}\sum_{jk}\mathbf{B}^{jk}\partial_{j}\mathbf{B}^{ik}dt, \\
[\mathbf{B}\star d\mathbf{w}]^{i} & =[\mathbf{B} \circledast d\mathbf{w}]^{i}
+\sum_{jk}\mathbf{B}^{jk}\partial_{j}\mathbf{B}^{ik}dt.
\end{align}
Thus we can conclude as follows. When we have the Stratonovich-Fisk SDE, 
\begin{align}
d\mathbf{x}=\mathbf{A}dt+ \mathbf{B}\circ d\mathbf{w},
\end{align}
this is equivalent to the Ito SDE, 
\begin{align}
d\mathbf{x}^{i}=\{\mathbf{A}^{i}+\frac{1}{2}\sum_{jk}\mathbf{B}%
^{jk}\partial_{j}\mathbf{B}^{ik}\}dt+\mathbf{B} \circledast d\mathbf{w}.
\end{align}
When we have the H\"{a}nggi-Klimontovich SDE, 
\begin{align}
d\mathbf{x}=\mathbf{A}dt+ \mathbf{B}\star d\mathbf{w} ,
\end{align}
this is equivalent to the Ito SDE 
\begin{align}
d\mathbf{x}^{i}=\{\mathbf{A}^{i}+\sum_{jk}\mathbf{B}^{jk}\partial _{j}%
\mathbf{B}^{ik}\}dt+ \mathbf{B} \circledast d\mathbf{w}.
\end{align}


\end{document}